\documentclass[lettersize,journal]{IEEEtran}
\usepackage{amsmath,amsfonts}
\usepackage{algorithmic}
\usepackage{algorithm}
\usepackage{array}
\usepackage[caption=false,font=normalsize,labelfont=sf,textfont=sf]{subfig}
\usepackage{textcomp}
\usepackage{stfloats}
\usepackage{url}
\usepackage{verbatim}
\usepackage{graphicx}
\usepackage{cite}
\usepackage{amsmath,amssymb,amsfonts}
\usepackage{algorithmic}
\usepackage{graphicx,color}
\usepackage{textcomp}
\begin{document}

\title{Three-Dimensional Millimeter-Wave Imaging Using Active Incoherent Fourier Processing\\and Pulse Compression}

\author{Jorge R. Colon-Berrios,~\IEEEmembership{Graduate Student Member,~IEEE,} Jason M. Merlo,~\IEEEmembership{Graduate Student Member,~IEEE,} Jeffrey A. Nanzer,~\IEEEmembership{Senior Member,~IEEE}
        % <-this % stops a space
\thanks{Manuscript received 2025.}
\thanks{This paper is an expansion of a paper presented at the 2025 IEEE International Microwave Symposium in San Francisco, CA.}% <-this % stops a space
\thanks{The authors are with the Department of Electrical and Computer Engineering, Michigan State University, East Lansing, MI 48824 USA (email: colonbe1@msu.edu, merlojas@msu.edu, nanzer@msu.edu).}
%\thanks{Manuscript received 2025.}
}


%{0000--0000/00\$00.00~\copyright~2021 IEEE}

\maketitle

\begin{abstract}

We present a novel three-dimensional (3D) imaging approach that combines two-dimensional spatial Fourier-domain imaging techniques with traditional radar pulse compression to recover both cross-range and down-range scene information. The imaging system employs four transmitters, three of which emit spatially and temporally incoherent noise signals, while the fourth transmits a known linear frequency modulated (LFM) pulsed signal. The spatial incoherence of the noise signals enables sampling of the 2D spatial Fourier spectrum of the scene from which two-dimensional cross-range (azimuth and elevation) images can be formed via interferometric processing. Simultaneously, the LFM signal enables high-resolution downrange imaging through matched filtering. The received signals consist of a superposition of the noise sources and the known pulse allowing for joint recovery of all three dimensions. We describe the system architecture and waveform design, and demonstrate the imaging technique using both simulations with a linear array and experimental data from a 38~GHz active incoherent millimeter-wave imaging system with 23-element randomized array. Results show the reconstruction of targets in three dimensions.
\end{abstract}


\section{Introduction}

Imaging at millimeter-wave frequencies offers unique advantages for sensing through visually or physically obstructed environments. These systems are very useful for tasks such as concealed object detection, gesture recognition, and atmospheric observation. The small wavelengths involved enable high spatial resolution, while the signals themselves can propagate through materials like clothing, fog, smoke, and certain building materials~\cite{currie1987principles,nanzer2012microwave,7836336}. Some systems operate passively by collecting naturally emitted thermal energy from objects in the scene, which requires highly sensitive detection electronics to measure faint thermally-generated radiation~\cite{1492659}. Other approaches improve signal quality by actively sending waveforms and capturing the reflected signals, which allows for greater control over the sensing process and reduces dependence on receiver sensitivity also known as active systems~\cite{Hunt310,10.1117/12.488198}. Signal processing strategies also differ across systems. Techniques that preserve and utilize phase information tend to achieve precise localization and imaging, but require strict synchronization and detailed knowledge of the transmitted waveform. In contrast, incoherent methods rely on signal intensity rather than phase and can function without such constraints. A widely used example involves spatially distributed sensors that analyze correlations in the received energy to obtain information from the scene structure~\cite{Thompson2001}. However, while these methods can capture angular features, they typically lack the ability to resolve target distance.

Active imaging systems address the low signal power in passive millimeter-wave sensing by transmitting energy into the scene, enhancing the backscattered return and allowing for the use of conventional receiver architectures achieving high angular resolution. They also require densely populated apertures such as phased arrays, focal plane arrays, or mechanically scanned reflectors which can significantly increase hardware cost and overall system complexity~\cite{Skolnik2008,Stimson2014}. Fourier domain interferometric imaging provides a more hardware efficient alternative by using sparse antenna arrays that sample the spatial frequency domain and do not require a fully filled aperture~\cite{Thompson2001,Mailloux2005,Diebold:21}. Related efforts have also demonstrated compact 3D imaging systems using frequency-scanning antennas driven by FMCW signals, avoiding complex phased-array architectures while still achieving range and angular resolution~\cite{7485168}. Photonics-based Inverse Synthetic Aperture Radar (ISAR) systems have similarly demonstrated high-resolution 3D imaging through de-chirp processing and LFM waveform design~\cite{8733807}. However, these systems rely on target motion to synthesize large apertures and typically require specialized photonic hardware. In contrast, our approach operates with a stationary sparse antenna array, using a hybrid of active waveform transmission and incoherent interferometric processing. This enables accurate 3D scene reconstruction using conventional millimeter-wave hardware without relying on object motion or fully coherent system design.

Three-dimensional imaging is critical in applications requiring spatial localization across azimuth, elevation, and range dimensions, such as security screening~\cite{942570} and autonomous navigation~\cite{8100174}. Traditional Fourier domain interferometric imaging techniques offer efficient hardware implementations by leveraging sparse arrays for high angular resolution, but they inherently lack precise downrange resolution due to their use of temporally incoherent noise illumination~\cite{Vakalis2020,Diebold2021}. Recent work has addressed this limitation by introducing pulsed noise sources, which allow for coarse range discrimination by encoding limited temporal information into the scene response~\cite{Vakalis2023TMTT}. In this work, we build upon these techniques by incorporating a linear frequency modulated (LFM) pulse into the active incoherent imaging system. The use of LFM enables pulse compression through matched filtering, improving downrange resolution while preserving the advantages of incoherent cross-range imaging. Compared to prior pulsed noise-based approaches, the LFM waveform provides enhanced range localization without requiring fully coherent transmission or dense aperture architectures.

In this article, we present a millimeter-wave imaging technique that integrates noise based interferometric processing with pulse compression using an LFM signal. Building on our previous work in~\cite{colon2025ims}, where scene reconstruction was demonstrated using pseudo-random noise pulses and pulse compression, this paper introduces the use of an LFM waveform to enhance downrange resolution, and conducts extensive proof-of-concept experimentation. We examine the system downrange capabilities in greater detail, analyzing how pulse bandwidth and waveform design affect the spatial response. The introduction of structured waveforms enables the system to resolve multiple targets within the same scene and also track their motion across the azimuth-elevation dimension. This hybrid approach offers a significant improvement over prior incoherent methods, achieving three-dimensional imaging with increased accuracy.
\begin{figure*}[t!]
	\centering
	\includegraphics[width=1\textwidth]{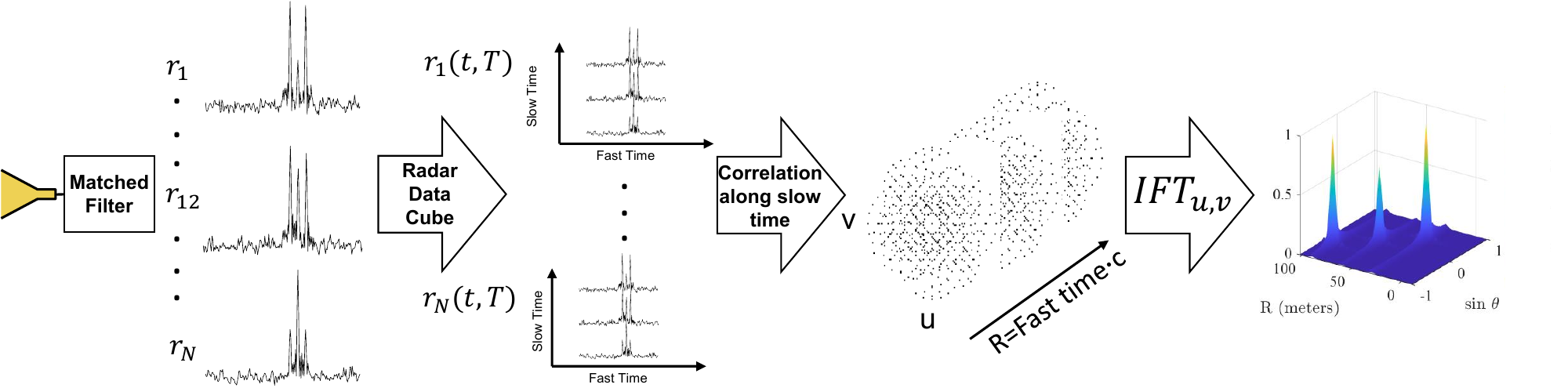}

	\hfil
	\caption{The proposed imaging approach is based on a combination of matched filtering and Fourier domain interferometric imaging. The array transmits noise signals and a known wavefom (a linear frequency modulated waveform in this work), which are reflected off the scene and captured at multiple receiving elements. Each channel undergoes pulse compression by correlating the received signal with the known transmitted waveform. The pulse-compressed data is then organized into a radar data cube, with axes representing fast time (range), slow time (pulses), and antenna element (position). Cross-correlation is performed along the slow time across pulses to extract cross-range information via Fourier domain interferometric processing. Finally, an image reconstruction is performed obtaining an image showing down range information of the target and azimuth location of the target.}
	\label{OV1}
\end{figure*}
\section{Active Fourier Domain Imaging}

%%%%%%%%%%%%%%%%%%%%%%%%%%%%%%%%%%%%%%%%%%%%%%%%%%%%%%%%%%%%%%%%%%%%%%%%%%%%%

Interferometric imaging makes use of spatial diversity across antenna arrays to reconstruct the angular structure of a scene by measuring the coherence of signals received between separate antennas. In this approach, the spatial distribution of energy in the scene is reconstructed from spatial frequency samples measured via a set of antenna baselines.
Let the scene intensity be given by $I(\alpha, \beta)$, where the direction cosines $\alpha = \sin\theta\cos\phi$ and $\beta = \sin\theta\sin\phi$ represent the angular coordinates in the azimuth and elevation planes. When signals received at different antenna locations are cross-correlated, the system measures the \textit{visibility} $V(u, v)$, which is the two-dimensional Fourier transform of the scene intensity:
\begin{equation}\label{eq.vis}
	V(u,v) = \iint\limits_{-\infty}^{+\infty} I(\alpha,\beta) \, e^{j2\pi(u\alpha + v\beta)} \, d\alpha \, d\beta\mathrm{.}
\end{equation}
Here, $u$ and $v$ denote the spatial frequency coordinates, determined by the geometry of the antenna pair: $u = D_x/\lambda$, $v = D_y/\lambda$, where $D_x$ and $D_y$ are the baseline distances between two antennas in the $x$ and $y$ directions, respectively.

Each antenna pair contributes a visibility sample at a point $(u_n, v_m)$, and the collection of these samples forms a sampling function:
\begin{equation}\label{eq.sf}
	S(u,v) = \sum_{n=1}^{N} \sum_{m=1}^{M} \delta(u - u_n) \, \delta(v - v_m)\mathrm{,}
\end{equation}
where $N \times M$ is the total number of visibility samples and $\delta(\cdot)$ is the Dirac delta function. The system observes the product of the sampling function and the visibility function, and the image is reconstructed by applying an inverse Fourier transform:
\begin{equation}\label{eq.recon}
	I_r(\alpha,\beta) = \iint\limits_{-\infty}^{+\infty} V(u,v) \, S(u,v) \, e^{-j2\pi(u\alpha + v\beta)} \, du \, dv\mathrm{.}
\end{equation}

In practice, with discrete sampling, the reconstructed image becomes a summation:
\begin{equation}\label{eq.ir}
	I_r(\alpha,\beta) = \sum_{n=1}^{N} \sum_{m=1}^{M} V(u_n, v_m) \, e^{-j2\pi(u_n \alpha + v_m \beta)}\mathrm{.}
\end{equation}
By taking an inverse Fourier transform on the sampling function we obtain a point spread function (PSF),
\begin{equation} \label{PSF}
	\mathrm{PSF}=F^{-1} \left \{ {S(u,v)}  \right \}
\end{equation}
which gives an idea of the quality with which the imaging system will reconstruct images. The angular resolution of an interferometric imager in the azimuth and elevation planes can be given in terms of the the half-power beamwidth $\theta_{\mathrm{HPBW}}$ of the sinc-squared response from the largest baselines \textcolor{black}{$D_x$ and $D_y$} in the horizontal and vertical axes of the array $x$ and $y$, defined by
\begin{equation}\label{res2}
	\Delta \theta_{\alpha,\beta} \approx  \theta^{(\alpha,\beta)}_{ \mathrm{HPBW}} \approx 0.88 \frac{\lambda}{D_{x,y}}. 
\end{equation} 
The field of view (FOV) is given by the smallest  \textcolor{black}{inter-}element spacings $d_x$ and $d_y$ across the horizontal and vertical axes can be expressed for the two direction cosines $\alpha$ and $\beta$ as

\begin{equation}\label{fov}
	\mathrm{FOV}_{\frac{\alpha}{2},\frac{\beta}{2}} = \frac{\lambda}{2 \cdot d_{x,y}}.
\end{equation} 

This two dimensional imaging framework forms the basis of Fourier domain interferometry, commonly applied in systems where the receive signal is spatially and temporally incoherent, consistent with the Van Cittert–Zernike theorem~\cite{Thompson2001}. Thermal radiation inherently meets these requirements, but its low power at millimeter-wave frequencies demands the use of high-gain, low-noise receivers~\cite{yujiri2003passive}. By implementing noise transmitters, which emulate the incoherence of natural thermal sources while offering greater power levels~\cite{8458190}. By placing the transmitters at diverse spatial locations ensures incoherent illumination across the array.
 
A limitation of this approach is the lack of downrange resolution information, since temporal information is not preserved due to the incoherent nature of the signals. To overcome this challenge, the technique presented in this work integrates known pulsed waveforms, specifically LFM signals, into the active incoherent imaging framework. By applying matched filtering and pulse compression, the time delay information is recovered from the reflected signals, enabling estimation of downrange distance in addition to the cross-range structure. This hybrid technique retains the spatial diversity of interferometric imaging while providing range resolution comparable to that of pulse radar systems. The result is a fully three-dimensional millimeter-wave imaging system capable of high resolution scene reconstruction in both angular and range dimensions.

\section{Down-Range Target Estimation Via Pulse Compression}

Illumination of the scene requires at least three noise transmitters pairwise oriented with extension in the azimuth and elevation directions to generate the required spatial and temporal incoherence~\cite{9127123}. In particular, at least two elements must extend beyond the spacing of the receiving array in both dimensions to ensure that the combination of the transmitted noise signals yields a signal that is spatially incoherent at a resolution narrower than that of the receiving array. To accommodate down-range measurement, we add to these three noise transmitters a fourth transmitter emitting an LFM signal. The signal scattered from the scene back to the receiving array can then process the received LFM along with the 2D cross-range information to generate 3D measurements of the scene.

%The LFM waveform is a widely used radar signal due to its high time-bandwidth product, which enables improvements in range resolution after matched filtering.
The LFM waveform is widely used in radar systems due to its favorable ambiguity function, which offers high range resolution after matched filtering. In this work we implemented an LFM with a linear frequency chirp sweeping across a bandwidth of 100~MHz over a pulse duration of 10~\textmu s. This results in a time-bandwidth product of 1000, which indicates high potential range resolution after pulse compression. The waveform was sampled with a time step of 10~\text{ns}. The signal was generated at baseband using an arbitrary waveform generator and then to input to a upconverter for transmission. The received signal was downconverted and matched filtered at baseband after digitization. The full waveform consisted of multiple pulses transmitted in a regularly spaced pattern. Ten pulses were repeated every 50~\textmu s for a total waveform duration of 500~\textmu s.
The LFM signal can be expressed as
 \begin{equation}
 s(t) = e^{j 2\pi \left( f_c t + \frac{B}{2T} t^2 \right)}, \quad 0 \leq t \leq T,
 \end{equation}
 where \( f_c \) is the carrier frequency, \( B \) is the bandwidth, and \( T \) is the pulse duration. In this case, the center frequency is \( f_c = 38~\text{GHz} \), the bandwidth is \( B = 100\,\text{MHz} \), and the pulse width is \( T = 10~\mu\text{s} \).
  One of the key advantages of the LFM waveform is its ability to achieve high down range resolution through pulse compression. The theoretical downrange resolution of the system is given by 
 \begin{equation}
 \Delta R = \frac{c}{2B},
 \end{equation}
where \( c \) is the speed of light. In this work we have a range resolution of 1.5 m.
%However with certain phase conditions along with high signal-to-noise ratio (SNR), it is possible to distinguish targets closer than this theoretical limit due to the sharpness of the matched filter response.

\begin{figure}[t!]
	\centering
	\includegraphics[width=0.24\textwidth]{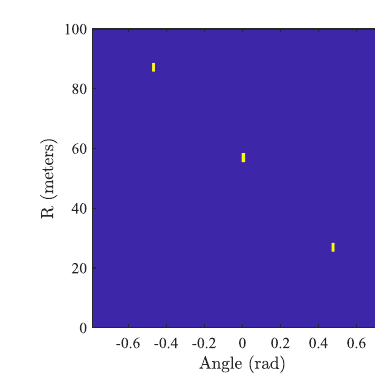}
	\includegraphics[width=0.24\textwidth]{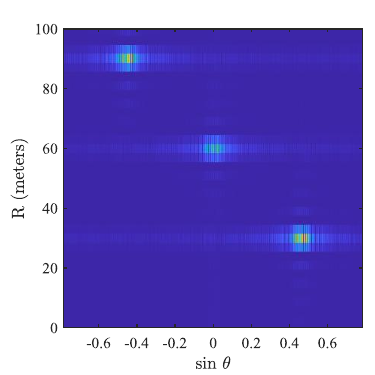}
	
	(a)
	\hfil
	
	\includegraphics[width=0.24\textwidth]{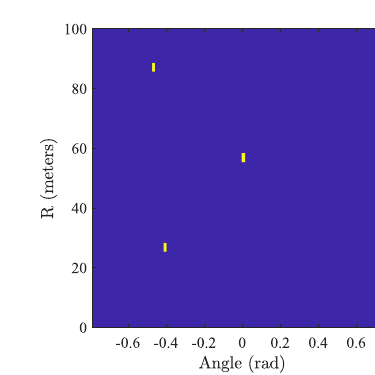}
	\includegraphics[width=0.24\textwidth]{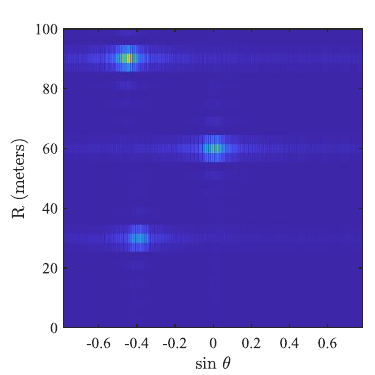}
	
	(b)
	
	\includegraphics[width=0.24\textwidth]{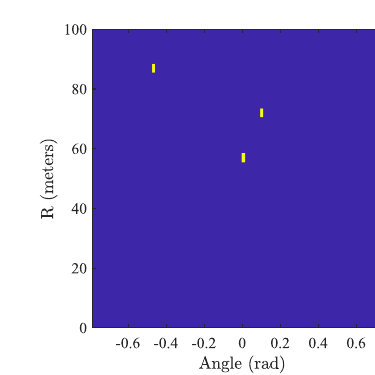}
	\includegraphics[width=0.24\textwidth]{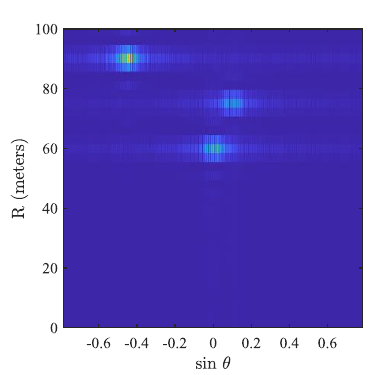}
	
	(c)
	\caption{Simulation of a linear array of 24 antenna elements spaced at half wavelength spacing. Three point targets were changed of placement to show the proposed method follows the target positions in downrange and crossrange.}
	\label{Sim}
\end{figure}

 %To overcome this limitation, in this work we use a different transmission method where one of the transmitters emits a pseudo-random noise pulse. The scattered signals are received by the array and formed into a 3D data cube consisting of information in the fast time (time of flight), slow-time (pulse to pulse), and channel (antenna) domains. The received signals can then be processed via matched filter designed with the known pseudo-random noise signal. 
%The technique consists on transmitting three signals that are long noise waveforms, while adding one noise pulse illumination. These signals bounce from the target and are received on the 24 receiver array. The matrix with the receive data is reshaped into a 3D data cube that consists of slow time, fast time, the last is the antennas channels. In this technique, we obtain our received signals and match filter them against the pulsed noise waveform.

The signals scattered from the scene thus contains the superposition of three noise signals and the LFM waveform. 
%Each receiver contains the superposition of the noise signals and the known LFM pulse. 
%The received signals are filtered using the LFM pulse train as the reference signal through
Cross-range images can be formed using the 2D inverse Fourier transform as described above, and down-range information can be obtained by processing the LFM on the received signals via a matched filter
%\begin{equation}\label{match}
%	C_i[k] = \sum_{n=-\infty}^\infty r[n] \cdot s_i[n + k], \quad i = 1, 2, \ldots, N
%\end{equation}
\begin{equation}\label{match}
	C_i[k] = \sum_{n=-\infty}^\infty s_i[n] \cdot h^*[n - k], \quad i = 1, 2, \ldots, N
\end{equation}
where $h[n]$ is the reference waveform, $k$ is the discrete time lag, and $h^*[n - k]$ denotes the time-reversed and conjugated filter and $s_i$ is the signal from the $i$th channel~\cite{richards2014fundamentals}. The responses are then stored in columns that represent the receiving antenna channel, after which Fourier domain processing is implemented for cross-range imaging. Note that although the LFM generates a coherent signal across all the receivers, the three noise transmitters still provide spatial and temporal incoherence to support cross-range Fourier domain imaging.
%where $k$ is the discrete time lag, $r$ is the reference signal, 

The overall signal processing approach to generated 3D measurements is summarized in Fig.~\ref{OV1}. The received signals are first processed using the matched filter on each channel in fast time, after which the data is added to the radar data cube that consists of the radar channels and the pulse-to-pulse measurements (slow time). Cross correlation is performed across slow time, yielding the Fourier domain interferometric response that generates the azimuth and elevation cross-range image. The fast time response is the pulse compressed matched filter output, yielding the down-range information.

\section{Simulated Down-Range \slash Cross-Range Imaging Results}

 The approach was demonstrated in simulation utilizing a linear array of 24 antenna elements spaced at half the wavelength spacing. The carrier frequency was 38~GHz for a wavelength of 7.9~mm. The simulation modeled transmission and reception of signals over a 500~\textmu s interval, scanning across angles from $-45$~\textdegree~to $+45$~\textdegree. Among the transmitters, one emitted the LFM waveform while the other three continuously transmitted noise signals. The radar scene was composed of three point targets positioned at various ranges and angles. 
%The signals received at the antenna array were modeled as the superposition of the reflected signals from all transmitters that is, the reflected signals of the LFM pulse combined with reflections of the three continuous noise signals. 

The received signals were modeled as the superposition of the four signals with appropriate delays based on the locations of the transmitters, receivers, and scattering targets. 
%include a time delay and phase shift based on target positions and antenna array baseline. 
The processing of the signals is shown in Fig.~\ref{OV1}. Once the signals are received they are matched filtered in fast time (intrapulse) with the known LFM waveform to extract the range information. The matched filtered output is then reshaped into a radar cube where cross-correlation is performed along slow time (interpulse) to extract target angular information. The resulting image reconstruction consists of an azimuth angle and range image, shown in Fig.~\ref{Sim}. Three different cases were tested by placing three targets at varying ranges and azimuth positions. One of the targets was moved across different range and azimuth locations, and in each case, all three targets were successfully detected at their expected positions. Compared to the original configuration, the detections remained consistent.   %using a 24-element uniform linear receive array along with three continuous-wave noise transmitters and one pulsed noise transmitter.

\begin{figure}[t!]
	\centering
	\includegraphics[width=0.45\textwidth]{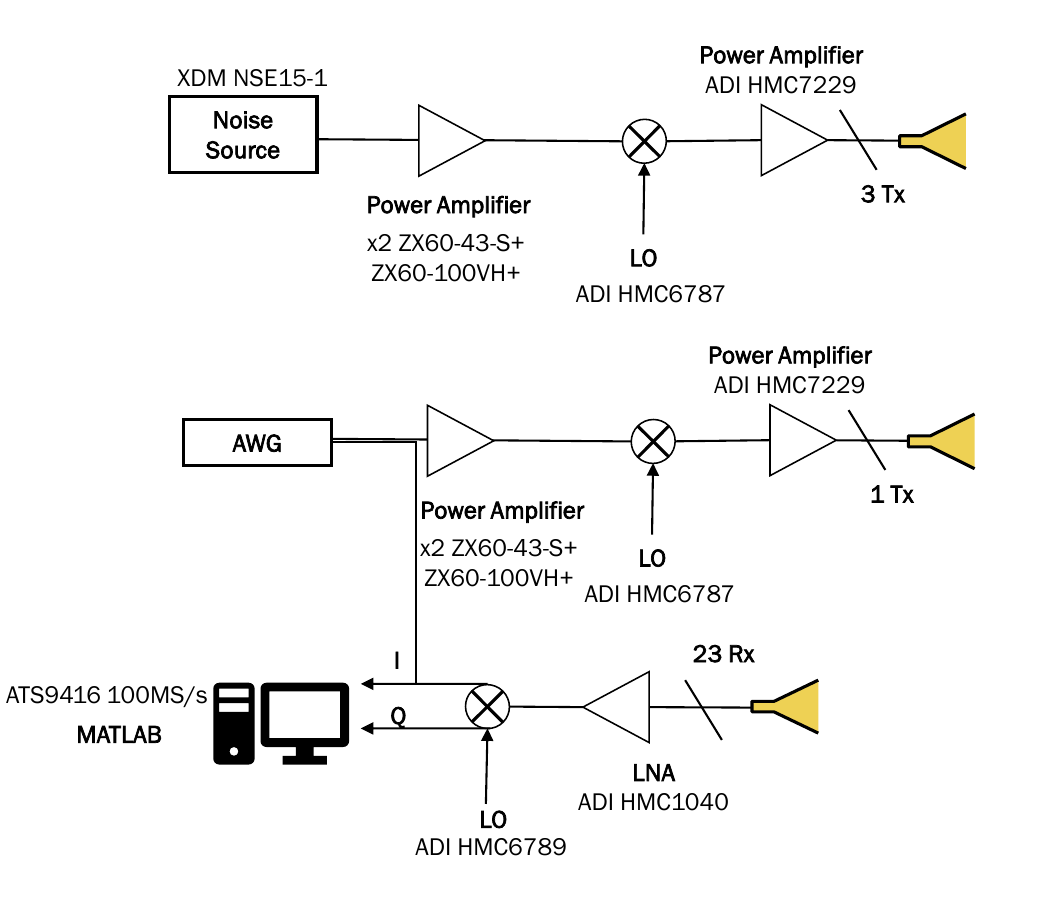}
	
	(a) 
	\hfil
	
	\includegraphics[width=0.45\textwidth]{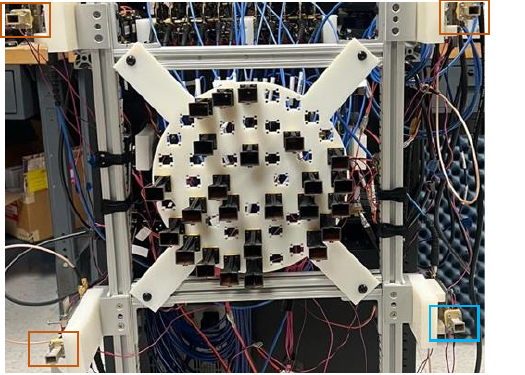}
	
	(b)
	\hfil
	
	\includegraphics[width=0.4\textwidth]{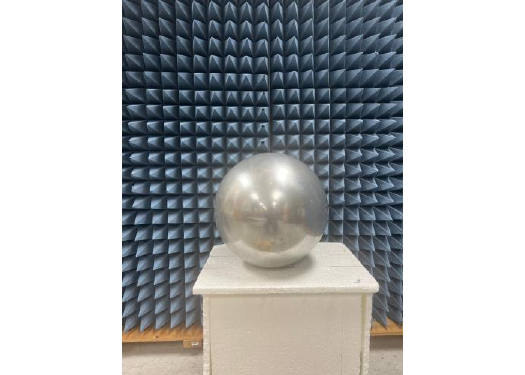}
	
	(c)
	
	\hfil
	\caption{(a) System schematic. The imager consists of three transmitters emitting a continuous noise signal and one transmitter emitting the LFM pulse train. The LFM signal is also connected to one of the 48 I/Q channels. This channel is used to perform match filtering for down range processing. (b) Image of the 38~GHz imaging system. The transmitters shown in the corners of the image consisted of three continuous-wave noise transmitters (in red boxes) and one pulsed noise transmitter (in the blue box). The receiving elements were placed in a randomized pattern to reduce redundant baselines and increase the number of unique visibility samples. (c) Image of the target used in the experiments, consisting of a large sphere that is moved to different locations in the cross-range.}
	\label{System}
\end{figure}

\begin{figure}[t!]
	\centering
	\includegraphics[width=0.24\textwidth]{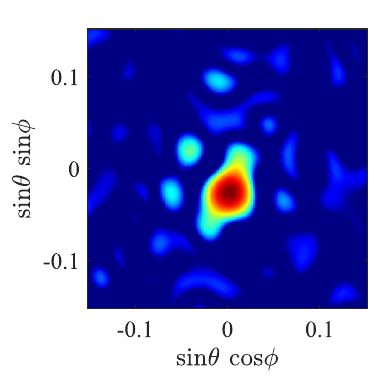}
	\includegraphics[width=0.24\textwidth]{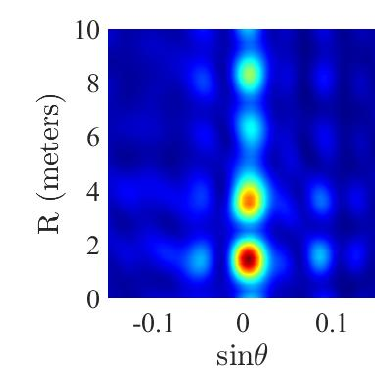}
	
	(a)
	
	\includegraphics[width=0.24\textwidth]{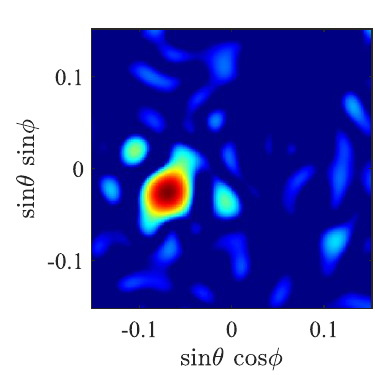}	
	\includegraphics[width=0.24\textwidth]{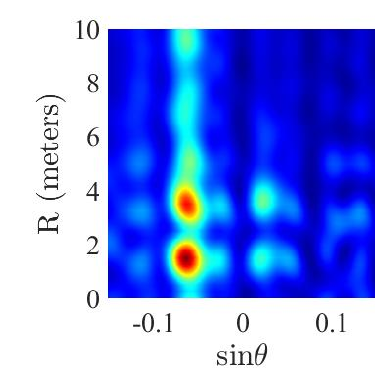}
	
	(b)
	
	\includegraphics[width=0.24\textwidth]{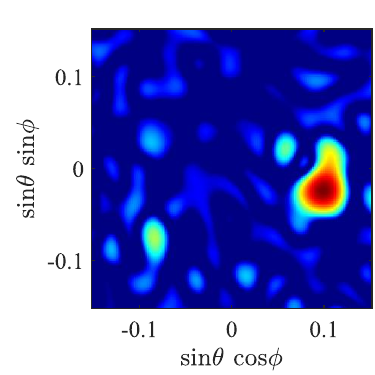}	
	\includegraphics[width=0.24\textwidth]{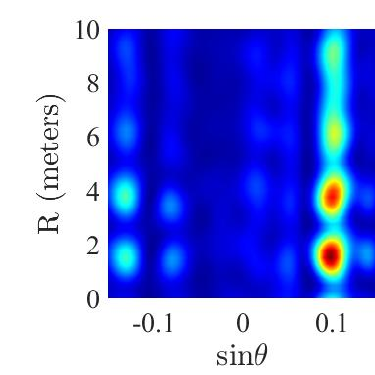}
	
	(c)
	\caption{Experimental results showing a sphere at 2 m (a) Center of FOV (b) Left of the FOV (c) Right of the FOV. Three dimensions of information are seen as the target is prominent in cross range and down range correctly.}
	\label{Results}
\end{figure}

\begin{figure}[t!]
	\centering
	\includegraphics[width=0.24\textwidth]{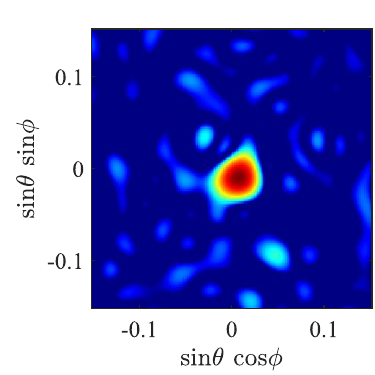}
	\includegraphics[width=0.24\textwidth]{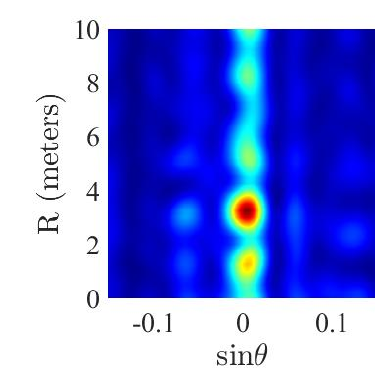}
	
	(a)
	\hfil
	
	\includegraphics[width=0.24\textwidth]{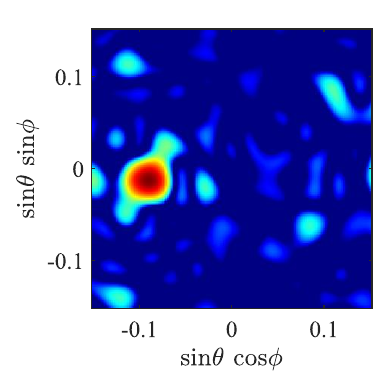}	
	\includegraphics[width=0.24\textwidth]{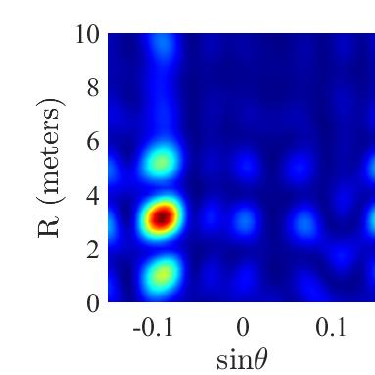}
	
	(b)
	
	\includegraphics[width=0.24\textwidth]{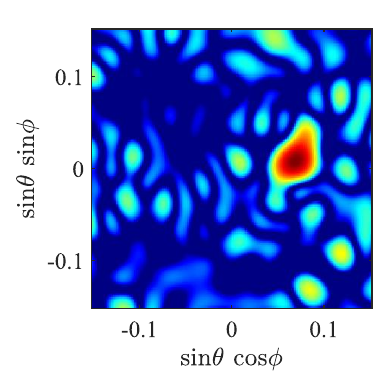}	
	\includegraphics[width=0.24\textwidth]{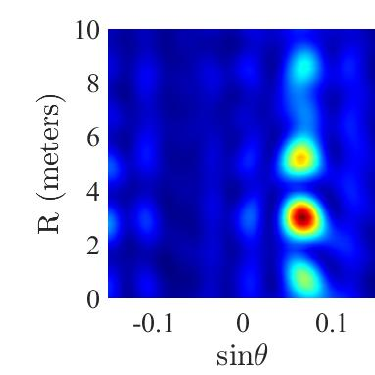}
	
	(c)
	\caption{Experimental results showing a sphere at 3 m (a) Center of FOV (b) Left of the FOV (c) Right of the FOV. At each case the sphere was detected correctly at 3 m. In the case (c) the sphere was elevated to see a different elevation and the downrange was detected correctly.  }
	\label{Results2}
\end{figure}

\begin{figure}[t!]
	\centering
	\includegraphics[width=0.4\textwidth]{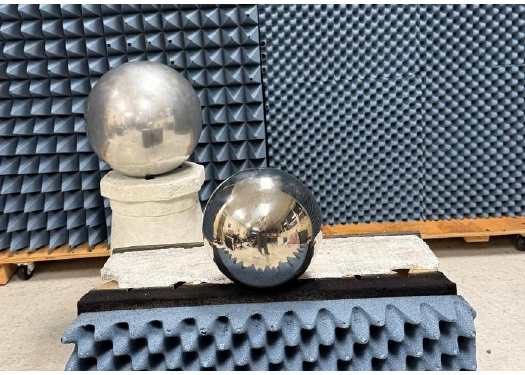}
	
	(a)
	\hfil
	
	\includegraphics[width=0.4\textwidth]{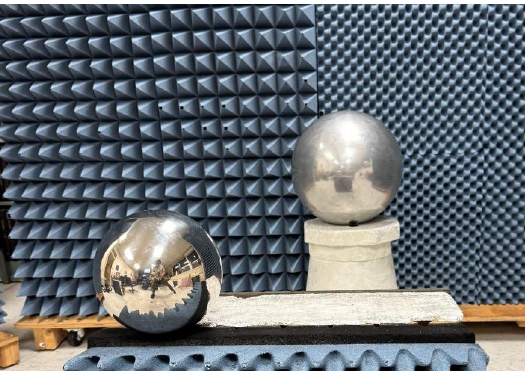}
	
	(b)
	\hfil
	
	\caption{Measurement setup for two targets at different ranges of 2~m and 3~m. (a) Sphere with a diameter of 30~cm and RCS of -11.5~dBsm at 3~m to the right of the FOV and a sphere of 18~cm diameter and RCS of -15.94~dBsm placed at 2~m and to the left of the FOV. (b) The scene is similar changing the placement of the spheres to 3~m to the left and 2~m to the right.      }
	\label{Optical2Targ}
\end{figure}

\begin{figure}[t!]
	\centering
	\includegraphics[width=0.24\textwidth]{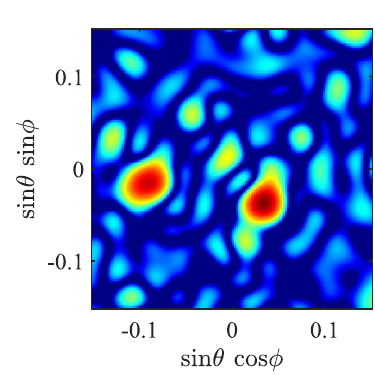}
	\includegraphics[width=0.24\textwidth]{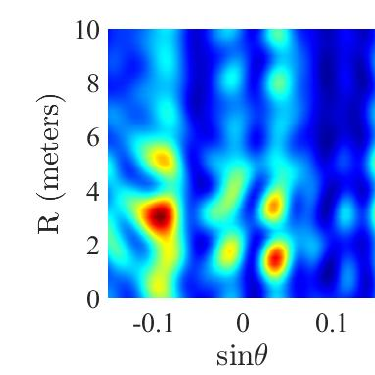}
	
	(a)
	
	\includegraphics[width=0.24\textwidth]{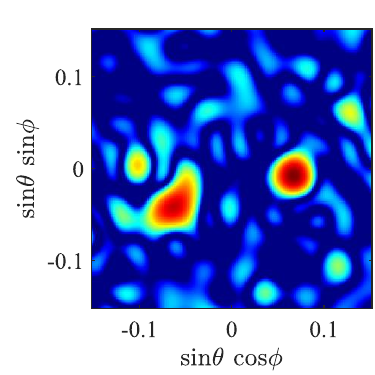}	
	\includegraphics[width=0.24\textwidth]{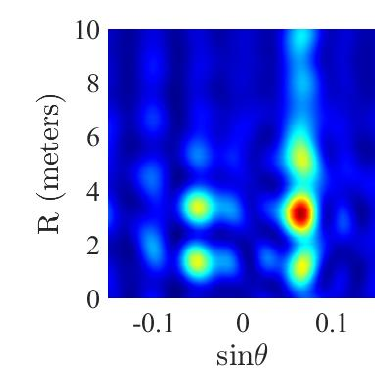}
	
	(b)
		
	\caption{Experimental results showing 2 targets at 3~m and 2~m  (a) The right target is at 3~m while left target at 2~m (b) The right target is at 2~m while left target at 3~m  }
	\label{Results3}
\end{figure}

\section{Experimental Verification}

\subsection{Experimental Setup}

Experimental measurements were conducted in a semi-anechoic environment using a metal sphere with a diameter of 30~cm. The sphere was placed at distances of 2 m and 3 m from the imaging system. In addition, the sphere was moved in angle to the right and left in the FOV. We also measured two targets present in the scene at different ranges of 2~m and 3~m. For the case of two targets a sphere of 18~cm diameter was utilized in addition with the original target. The schematic of the 3D AIM system and an image of the system is shown in Fig.~\ref{System}. 

The array combined 24 receive antenna elements in a randomized array formation with four transmitters, three of which were transmitting noise and one transmitter that illuminated the scene with a LFM pulse. 24 3D-printed standard gain horn antennas with approximately 15~dBi gain that were sputter-coated in copper were used on the receiving array, while metal 10~dBi standard gain horn antennas were used on the transmitters. The transmitters were placed at a wider baseline than the receivers to ensure that the signals incident on the scene were spatially uncorrelated at a spatial resolution finer than that of the receiver array. The noise signals were generated using calibrated noise sources (XDM NSE15-1), the signals from which were amplified at baseband using Mini-Circuits ZX60-43-S+ power amplifiers and then upconverted using Analog Devices (ADI) HMC6787ALC5A upconverters with a 19~GHz local oscillator (LO) input to a frequency doubler integrated onto the upconverters. The carrier frequency was generated by a Keysight N5183A signal generator. The LFM pulse was implemented by modulating the intermidiate fre quency(IF) on the upconverter with a 100~MHz LFM pulse generated using a Keysight Arbitrary Waveform Generator (AWG), which was then amplified at baseband and upconverted using the same devices as the noise signals.
%The LFM coded follows very closely the simulated signal with 50~MHz over a pulse duration of 10~\textmu s. 
The LFM waveform consisted of ten 50~\textmu s pulses and had a total duration of 500~\textmu s.   %The sphere was moved between positions of 2.74 meters and 3.36 meters. The PRF of this signal was 5~MHz.  
%Each transmitter used a separate calibrated baseband noise source, the signals from which were first amplified using three amplifiers two ZX60-43-S+ and one  ZX60-100VH+ in cascade and then upconverted to the millimeter-wave carrier frequencies using Analog Devices (ADI) HMC6787 upconverters. 
Before being transmmitted by the horn antennas, all four millimeter-wave signals were amplified using ADI HMC7229 power amplifiers. 

The signals were transmitted towards the scene, reflected back, and captured by the receiving array. 
One of the receiver channels was used as a loop-back receiver to sample the transmitted LFM signal, thus only 23 signals were captured in the receiving array.
After each 15~dBi horn antenna, the received signals 
%The upconverted noise signals and LFM pulse were transmitted to the scene and scattered back towards the receiving array. The signals captured at the receiving array 
were initially amplified using ADI HMC1040 low-noise amplifiers and then quadrature downconverted using ADI HMC6789 downconverters fed with the 19~GHz LO used on the transmitter. 
After downconversion of the 23 received signals, the 46 I/Q baseband signals were digitized by three ATS9416 samplers that were hosted in a computer. The sample rate on each channel was 100~MSa/s. All signal processing and image reconstruction was implemented in a host computer using MATLAB. 
The transmitted LFM signal was coupled to the baseband input of one of the receiving channels directly, and was used as a reference signal for matched filtering.

\subsection{Measurement Results}

Fig.~\ref{Results} shows the results of detection of the large spherical metal target at at distance of 2~m away from the system. The figures show the azimuth--elevation image, formed via 2D Fourier processing, in the left column, and the cross-range--down-range image, formed using 1D Fourier domain processing in azimuth and pulse compression in down range, in the right column. The images in both columns show a clear response of the sphere as it moves from center to the left and then to the right in the field of view. The down-range--cross-range image exhibits appreciable down-range sidelobes due to the matched filter processing; these are slightly larger than expected for the LFM waveform, which is partly due to signal saturation because of the proximity of the target to the system; subsequent measurements at larger distances resulted in reduced sidelobes. Nonetheless, the principal response in each image accurately tracks the distance and angle of the target, demonstrating the efficacy of the proposed 3D imaging technique.

Results of the same target at a distance of 3~m are shown in Fig.~\ref{Results2}. In these measurements the principal responses in the images again track the range and angle of the target. Because the SNR is slightly lower at longer ranges, the azimuth--elevation image displays a few more noise artifacts, however the principal response remains strong. The down-range--cross-range image exhibits a clearer response than when the target was placed at 2~m, likely because of the reduction in signal saturation from the reduced signal strength. The range sidelobes are significantly reduced as well, and the leading sidelobe is also visible in the field of view.

We then conducted measurements of the two spheres at different ranges and angles. The measurement setup is shown in Fig.~\ref{Optical2Targ}. The spheres were placed at two angles with a down-range separation of 1~m, with each sphere at 2~m and 3~m distance from the system. The measurement results are in Fig.~\ref{Results3}, showing the azimuth-elevation images for the two measurements along with the down-range--cross-range images. The two targets are clear in each image and the prominent responses track the angles and ranges of the targets. Some strong sidelobe structure is evident, along with some artifacts present at angles between the two targets. This may be due to signal saturation from the near target, but also from multiple bounce scattering between the two reflecting spheres, which can be appreciable with such targets due to their omnidirectional scattering pattern. Nonetheless, the responses in the images appropriately track the locations of the targets, further demonstrating the efficacy of the proposed 3D Fourier domain imaging technique.

\section{Conclusion}

A new method for three-dimensional millimeter-wave imaging was presented and experimentally demonstrated. By combining active incoherent millimeter-wave imaging with traditional radar pulse compression, sufficient spatial and temporal incoherence can be obtained to create cross-range images in azimuth and elevation using interferometric Fourier imaging approaches while simultaneously illuminating the scene with a known waveform that can be processed via pulse compression for high resolution down-range processing. We demonstrated the technique in a 38~GHz active incoherent millimeter-wave imaging system, showing that the approach successfully tracks a single reflecting target and multiple targets simultaneously at different angles and ranges. The proof-of-concept results in this paper show the feasibility of the technique, and point to areas for further improvement, such as increased range and mitigating the impacts of range sidelobes. The demonstrated method may prove useful in a range of imaging applications, including security, home health, remote sensing, and human-computer interaction.

\bibliographystyle{IEEEtran}
\bibliography{IEEEabrv,reference_bib}

\newpage

% your own custom macro containing the $\backslash${\tt{includegraphics}} command to make things
%simpler here.)
 

%\bf{If you include a photo:}\vspace{-33pt}
%\begin{IEEEbiography}[{\includegraphics[width=1in,height=1.25in,clip,keepaspectratio]{fig1}}]{Michael Shell}
%Use $\backslash${\tt{begin\{IEEEbiography\}}} and then for the 1st argument use $\backslash${\tt{includegraphics}} to declare and link the author photo.
%Use the author name as the 3rd argument followed by the biography text.
%\end{IEEEbiography}



\vfill

\end{document}